# Tuning electromagnetic properties of $SrRuO_3$ epitaxial thin films via atomic control of cation vacancies

Sang A Lee[1], Seokjae Oh[1], Jegon Lee[1], Jae-Yeol Hwang[2], Jiwoong Kim[3], Sungkyun Park[3], Jong-Seong Bae[4], Tae Eun Hong[4], Suyoun Lee[5], Sung Wng Kim[2], Won Nam Kang[1], and Woo Seok Choi[1*]

[1]Department of Physics, Sungkyunkwan University, Suwon, 16419, Korea

[2]Department of Energy Sciences, Sungkyunkwan University, Suwon 16419, Korea

[3]Department of Physics, Pusan National University, Busan 46241, Korea

[4]Busan Center, Korea Basic Science Institute, Busan 46742, Korea

[5]Electronic Materials Research Center, Korea Institute of Science and Technology, Seoul 02792, Korea

[*]e-mail: choiws@skku.edu.

**Elemental defects in transition metal oxides is an important and intriguing subject that result in modifications in variety of physical properties including atomic and electronic structure, optical and magnetic properties. Understanding the formation of elemental vacancies and their influence on different physical properties is essential in studying the complex oxide thin films. In this study, we investigated the physical properties of epitaxial $SrRuO_3$ thin films by systematically manipulating cation and/or oxygen vacancies, via changing the oxygen partial pressure ($P(O_2)$) during the pulsed laser epitaxy (PLE) growth. Ru vacancies in the low-$P(O_2)$-grown $SrRuO_3$ thin films induce lattice expansion with the suppression of the ferromagnetic $T_C$ down to ~120 K. Sr vacancies also disturb the ferromagnetic ordering, even though Sr is not a magnetic element. Our results indicate that both *A* and *B* cation vacancies in an $ABO_3$ perovskite can be systematically engineered via PLE, and the structural, electrical, and magnetic properties can be tailored accordingly.**

## Introduction

Transition metal oxides give rise to various functional behaviors resulting from the strongly coupled charge, spin, lattice, and orbital degrees of freedom.[1,2] Defects such as elemental vacancies in complex oxides can modify the interplay among these degrees of freedom, providing further controllability of the crystalline lattice, electronic structure, and magnetic ordering.[3-5] Oxygen vacancies, which can be easily manipulated during the deposition, are among the most prominent examples that induce changes in electronic and optical properties of transition metal oxide thin films.[6-10] On the other hand, cation vacancies, e.g., A- and/or B-site vacancies in $ABO_3$ perovskites, can also be employed for the control of different physical properties such as lattice structure, ferroelectricity, ferromagnetism, and thermoelectricity.[11-14] To properly design and take advantage of desirable material characteristics of oxide thin films and heterostructures, it is essential to comprehend the formation and the roles of various elemental defects.

Itinerant ferromagnet $SrRuO_3$ (SRO) can be considered as a model system for studying the strong couplings among the degrees of freedom mentioned above and their modifications due to the controlled elemental vacancies. In bulk, SRO has an orthorhombic structure with a *Pbnm* space group, with the pseudocubic lattice parameter of $a_{pc}$ = 3.926 Å.[15] It exhibits a paramagnetic to ferromagnetic transition with Curie temperature ($T_C$) of ~160 K. Because the ferromagnetic metallic property results from the strong hybridization between the Ru 4*d* and O 2*p* orbitals, the oxygen and cation vacancies play key roles in determining the physical properties of SRO. Whereas most studies have focused on Ru and/or oxygen vacancies,[13,16-20] it is obvious that Sr vacancies also play an essential part in determining the fundamental physical behavior. For example, by merely changing the A-site cation from Sr to Ca, isomorphic $CaRuO_3$ does not show any long-range ferromagnetic order.

In this study, we investigated the strong correlation among the stoichiometry (both cation and oxygen), crystal lattice, and electronic/magnetic properties of SRO epitaxial thin films. The cation ratio was selectively controlled by changing the oxygen partial pressure ($P(O_2)$) during the pulsed laser epitaxy (PLE) growth. The cation vacancy alters the hybridization between the Ru 4*d* and O 2*p* orbitals, inducing systematic changes in the electric and magnetic properties of SRO epitaxial thin film. Based on our results, we suggest that Sr vacancies induced in SRO also suppress $T_C$ as Ru vacancies via subtle lattice distortion.

## Results and discussion

Figure 1 shows the x-ray diffraction (XRD) $\theta$-2$\theta$ scan (Fig. 1a) and reciprocal space mapping (RSM) (Fig. 1b-1e) of the (001)-oriented SRO epitaxial thin films grown at different $P(O_2)$s. The SRO thin films show a systematic increase in the pseudocubic $c_{pc}$-axis lattice constants with decreasing $P(O_2)$, whereas the in-plane lattices are coherently strained to the $SrTiO_3$ (STO) substrates. The larger $c_{pc}$-axis lattice constant of SRO thin film (3.926 Å) compared to the STO substrate (3.905 Å) imposes compressive strain (lattice mismatch of 0.54%) in the thin film. The full-width-at-half maximum value of the $\omega$-scan peaks around $(002)_{pc}$ SRO Bragg diffraction is ~0.02°, suggesting good crystallinity of the thin films. Also, the thin films show the well-defined Kiessig and Pendellösung fringes, indicating the sharp interface between the thin film and the substrate as well as the smooth film surface.[8] In addition to the expansion of the $c_{pc}$-axis lattice constant with decreasing $P(O_2)$, we observed the structural phase transition from orthorhombic to tetragonal in the off-axis XRD $\theta$-2$\theta$ scan of {204} STO Bragg reflections (See Supplementary Fig. S1). The structural phase transition has been understood in terms of oxygen vacancies, i.e., the octahedral tilt is suppressed owing to electron repulsion induced by oxygen vacancies.[16]

The nonstoichiometric SRO thin films were studied using x-ray photoemission spectroscopy (XPS) and Rutherford backscattering (RBS) for the characterization of elemental vacancies. The results indicate a strong correlation between the structure, i.e., unit cell volume (or $c_{pc}$-axis lattice constant) and cation nonstoichiometry. We also note that the type of dominant cation vacancy changed from Sr to Ru across the stoichiometric growth condition of $P(O_2)$ ~ 100 mTorr. Figure 2 shows the Sr/Ru ratio as a function of $P(O_2)$, based on the atomic concentrations obtained from both XPS and RBS.

The results consistently indicate that the Ru (Sr) vacancies prevail in the SRO thin films grown at $P(O_2)$ below (above) 100 mTorr, in addition to the oxygen vacancies which obviously increases with decreasing $P(O_2)$.[8] Here, we would like to emphasize that the oxygen vacancies are created along with the Ru vacancies in the SRO thin films. The formation of Ru-O vacancy in SRO thin film is quite different from the case of STO thin film, where the cation and oxygen vacancies can be separately controlled.[10] With an increasing Sr/Ru ratio, the unit cell volume also shows a monotonically increasing behavior. Indeed, the unit cell volume can be a measure of the Sr/Ru ratio, regardless of the value being greater or less than one. This is again in contrast to the case of STO, where the unit cell volume increases when the Sr/Ti ratio deviates from one.[10,21] The correlation between the cation stoichiometry and unit cell volume in SRO thin film can be understood in terms of subtle internal structural distortion induced by the Ru vacancy site.[18,21] It has been suggested that as Sr atoms substantially relax towards vacant Ru sites, the Ru-O-Ru bond angle flattens, giving a positive contribution to the expansion of the unit cell volume.[21] On the other hand, Sr vacancies in the SRO thin film do not expand the unit cell volume. It has been reported that Ru ions are relatively static against local movement within the SRO crystal, compared to Sr ions.[21] This static nature would leave Ru ions in their original positions even in the case of adjacent Sr vacancies. Instead of expanding the unit cell volume, we suggest that Sr vacancies could induce subtle internal structural distortion involving $RuO_6$ tilt angles, which will be discussed later.

The $P(O_2)$-dependent changes indicate that the Sr/Ru ratio can be systematically engineered by modifying the plume dynamics during the PLE process. In particular, the ablated elemental species undergo different scattering dynamics with the background gas depending on their mass, which determines the stoichiometry of the deposited thin film. A lighter element is more susceptible to the background gas, and becomes more deficient as the gas pressure increases. For example, for the growth of STO thin films, low $P(O_2)$ growth results in Sr (heavier element) vacancies.[10] In addition, recent studies show that the same trend can be observed for the growth of $BaTiO_3$, $CaTiO_3$, $La_{0.4}Ca_{0.6}MnO_3$, $EuAlO_3$, and $LiMn_2O_4$ thin films.[22-24] For SRO, Ru is heavier than Sr, so Ru vacancies prevail in the highly energetic plume condition (low $P(O_2)$ growth), consistent with other oxide thin film growth. The scattering of relatively lighter Sr depends strongly on the $P(O_2)$ level, much more than the heavier Ru, resulting in Sr vacancies in the high $P(O_2)$ growth condition. Such plume dynamics with the highly volatile nature of Ru enables systematic elemental control. Indeed, the fine engineering in cation stoichiometry using PLE allow us to conclude that the Sr/Ru ratio does not show a particularly large change across the orthorhombic-to-tetragonal phase transition that occurs at $P(O_2) \approx 20$ mTorr.[8] The gradual introduction of Ru vacancies builds up structural energy for the orthorhombic phase, and when the Sr/Ru ratio increases above ~1.3, the structure transforms into the tetragonal phase.

The elemental vacancies affect the hybridization between the Ru $4d$ and O $2p$ states significantly, leading to systematic modifications in the electric and magnetic properties of the SRO thin films. Figure 3a shows the temperature-dependent resistivity ($\rho(T)$) for the SRO epitaxial thin films grown at different $P(O_2)$s. All samples show metallic behavior as a function of temperature, with the presence of an anomaly in the temperature range of 120–150 K, which indicates the ferromagnetic transition temperature ($T_C$). The highest $T_C$ is ~150 K for the stoichiometric epitaxial SRO thin film ($P(O_2)$ = 100 mTorr), consistent with other SRO thin films grown on STO substrates.[13,25] As Ru vacancies are introduced in the thin film, $\rho(T)$ systematically increases over all temperature ranges examined, indicating that the reduced hybridization (or orbital overlap) between Ru $4d$ and O $2p$ diminishes the electric conductivity.[8] On the other hand, the SRO thin film with Sr vacancies shows the lowest resistivity at high temperatures. At low temperatures (< ~75 K), however, $\rho(T)$ is higher than the stoichiometric SRO thin film, manifesting the greater disorder owing to the cation vacancies.

To further investigate the transport properties, $\rho(T)$ of the SRO thin films was fitted using the relation $\rho(T) = \rho_0 + AT^{\alpha}$ ($\rho_0$ is residual resistivity, $A$ is a coefficient, and $\alpha$ is scaling parameter) for three different temperature regions (See Supplementary Fig. S4).[26] When $T > T_C$, $\rho(T)$ of all thin films shows temperature dependence with $\alpha = 0.5$, indicating bad metal behavior.[27,28] In the temperature range of 50–120 K, $\rho(T)$ could be well fitted with $\alpha = 1.5$, which suggests scattering of Fermi liquid (FL) electrons to the localized electrons with local bond length fluctuations below $T_C$.[19,29] At temperatures below 30 K, $\rho(T)$ depends on $T^2$, ($\alpha = 2$), indicating fully FL behavior. The $\rho_0$ values for the thin films grown at $P(O_2)$ = 10, 100, and 300 mTorr are 80.0, 39.1, and 61.5 μΩ cm, respectively. The large $\rho_0$ of the SRO thin films grown at $P(O_2)$ = 10 and 300 mTorr could be related to the elemental vacancies, which induce disorder in SRO thin film. The $A$ values for the SRO thin films grown at $P(O_2)$ = 10, 100, and 300 mTorr, are $7.8 \times 10^{-3}$, $1.5 \times 10^{-2}$, and $6.2 \times 10^{-3}$ μΩ cm $K^{-2}$, respectively. In principle, the $A$ value represents the effect of electron-magnon scattering,[17,29] and the stoichiometric SRO thin film expectedly shows the highest value, based on the strongest magnetic interaction. The largest $A$ value also indicates the largest slope in $\rho(T)$ arithmetically, so the largest residual resistivity ratio (*RRR*) is explained for the stoichiometric thin film.

Figure 3b shows magnetization as a function of temperature ($M(T)$) for the SRO thin films. By introducing elemental vacancies (either Sr or Ru), the ferromagnetic behavior becomes systematically suppressed. Consistent with the $\rho(T)$ behavior, $T_C$ decreases to ~120 K, as summarized in Fig. 3c. Figure 3d summarizes the *RRR* values (defined as $\rho$(300 K) / $\rho$(30 K)) and magnetization at 5 K. The $T_C$, $M$(5 K), and *RRR* values show the exact same $P(O_2)$-dependent trend, having a peak for the stoichiometric sample. This observation indicates that the electric transport and magnetic properties of SRO epitaxial thin films are closely related with the cation stoichiometry.[30] The Ru vacancies (for the samples grown at $P(O_2)$ < 100 mTorr) hamper the charge transfer across the Ru-O-Ru network, disrupting the ferromagnetic interaction, as well as the itinerant behavior. On the other hand, the SRO thin film with Sr vacancies (for the sample grown at $P(O_2)$ = 300 mTorr) also shows lower $T_C$, magnetization, and *RRR* value compared to the stoichiometric SRO thin film. This indicates that the Sr vacancies disturb the magnetic ordering, although Sr is not a magnetic element. Generally, the inter-site magnetic coupling between Ru atoms depends on Ru-O-Ru exchange path, i.e. Ru-O-Ru bond length and angle. Because the Sr-O covalency induces the local structural distortion in the perovskite structure, the inter-site exchange coupling in SRO thin films depends on the Sr vacancies as well.[31,32]

Finally, we performed annealing experiments to examine the effect of oxygen vacancies on the physical properties of SRO thin films. The result implies that Ru and oxygen vacancies are formed together in the SRO thin film, and is difficult to modify the concentration of each vacancy type independently. The SRO thin films were annealed at 700°C for 2 h in air. Usually, in such condition, oxygen can get into the oxide thin films with oxygen vacancies, e.g., for the case of STO. As shown in Fig. 4, the XRD results show that the crystal structure does not change owing to the annealing. In addition, we measured optical conductivity spectra ($\sigma_1(\omega)$) of the thin films before and after the annealing. All optical absorptions, i.e., Drude and four electronic transitions, labelled as α (Ru 4$d$ $t_{2g}$ → $t_{2g}$), A (O 2$p$ → Ru 4$d$ $t_{2g}$), $\beta$ (Ru 4$d$ $t_{2g}$ → $e_g$), and B (O 2$p$ → Ru 4$d$ $e_g$) at ~1.7, ~3.3, ~4.1, and ~6.2 eV, respectively, do not change after the annealing.[8,33] These results indicate that the crystal and electronic structures (and, accordingly, the electrical and magnetic properties as shown in the inset of Fig. 4b) of the SRO epitaxial thin films cannot be modified by oxygen annealing. This is in stark contrast to the case of STO. For STO, a drastic change in the oxygen vacancy concentration resulted in metal-to-insulator transition for a similar experiment, while the cation vacancy concentration did not change maintaining the crystal structure.[10] This rather surprising result might indicate that either (1) the oxygen vacancies cannot be compensated via simple annealing and/or (2) the change in the oxygen content does not alter the crystal and electronic structure significantly. Since oxygen vacancies are highly likely to modify the crystal and electronic structure, it seems that the oxygen vacancy compensation is more difficult for the case of SRO, possibly due to the fact that Ru and oxygen vacancies are strongly bound together within the crystal. In any case, we can conclude that the cation (Sr and Ru) vacancies are more controllable and influential in the SRO epitaxial thin films for the determination of the structural, electric, and magnetic properties.

## Summary

In summary, heteroepitaxial SRO (001) epitaxial thin films were grown coherently on STO substrates using PLE. Along with the unit cell volume, the Sr/Ru ratio could be systematically modified by controlling $P(O_2)$ during the PLE growth. From the stoichiometric condition of $P(O_2)$ = 100 mTorr, the lower $P(O_2)$ promotes Ru vacancies, which significantly affects the hybridization between the Ru 4*d* and O 2*p* orbitals. This results in the increased resistivity and the reduced ferromagnetic ordering of the epitaxial thin film. Sr vacancies, induced by growth in the higher $P(O_2)$, also disturbs the ferromagnetic property although Sr is a nonmagnetic element. We suggest that both Sr and Ru vacancies can be systematically engineered during the PLE growth, and play a crucial role in determining the crystal structure as well as electronic and magnetic properties. Our findings propose a way of exploiting elemental vacancies to realize functionality tailored complex transition metal oxide epitaxial thin films.

## Methods

**Thin film growth and structural characterization.** High-quality heteroepitaxial SRO thin films were grown on atomically flat STO single crystalline substrates using PLE at 700 °C.[8,34] Laser (248 nm; Lightmachinery, IPEX 864) fluence of 1.5 J/cm$^2$ and repetition rate of 2 Hz were used. To systematically control the elemental vacancies in SRO, the thin films were grown under various $P(O_2)$, ranging from $10^{-1}$ to $10^{-3}$ Torr. The thickness of the SRO thin films was ~30 ± 1 nm from x-ray reflectometry (XRR). The atomic structure, crystal orientations, and epitaxy relation of SRO thin films were characterized using x-ray diffraction (XRD) and reciprocal space mapping (RSM).

**X-ray photoemission spectroscopy (XPS) and Rutherford back scattering spectroscopy (RBS).** The chemical composition of the SRO thin films was studied using XPS and RBS at room temperature. For XPS, an Al-*Kα* monochromator x-ray source ($h\nu$ = 1486.6 eV) with a step size of 0.1 eV at a pass energy of 50.0 eV with a spot size of 400 $\mu$m was used. For RBS, an NEC 6SDH pelletron accelerator with energy of 3.0 MeV was used. The source gas was $He^{+2}$. The tilting angle was 5°. To distinguish between film and substrate, we prepared the SRO thin films grown on $LaAlO_3$ substrate at the same condition.

**Resistivity and magnetization measurements.** $\rho(T)$ was measured using a low-temperature closed-cycle refrigerator (CS202*I-KMX-1SS, Advanced Research System). The measurements were performed from 300 to 20 K, using van der Pauw electrode geometry with In electrodes and Au wires. $M(T)$ was measured using a Magnetic Property Measurement System (Quantum Design). The measurements were performed from 300 to 2 K under 10 Oe of the magnetic field along the in-plane direction of the thin film.

**Ellipsometry.** The optical properties of the SRO thin films were investigated using a spectroscopic ellipsometer (VASE, J. A. Woollam Co.) at room temperature. The optical spectra were obtained between 0.74 and 5.5 eV for the incident angles of 60°. A two-layer model (SRO thin film on STO substrate) was sufficient for obtaining physically reasonable spectroscopic dielectric functions of SRO, resembling those found in the literatures.

## Acknowledgements

This work was supported by Basic Science Research Programs through the National Research Foundation of Korea (NRF) (NRF-2017R1A2B4011083, NRF-2016R1A6A3A11934867 (S.A L.), and NRF-2015R1D1A1A01058672 (S.P.)). S.L. was supported by the Korea Institute of Science and Technology (KIST) through 2E25800. This work was also supported by IBS-R011-D1.

## Author contributions

S.A L, S.J.O, and J.L conducted the experiment and analyzed the results. J.Y.H and S.W.K performed structural analysis. J.W.K, J.S.B, and S.P performed XPS. T.E.H performed RBS analysis. S.L measured the electrical transport property. W.N.K performed magnetic property. S.A.L and W.S.C wrote the manuscript, and all the authors reviewed the manuscript. W.S.C initiated and supervised the research.

**Figure legends**

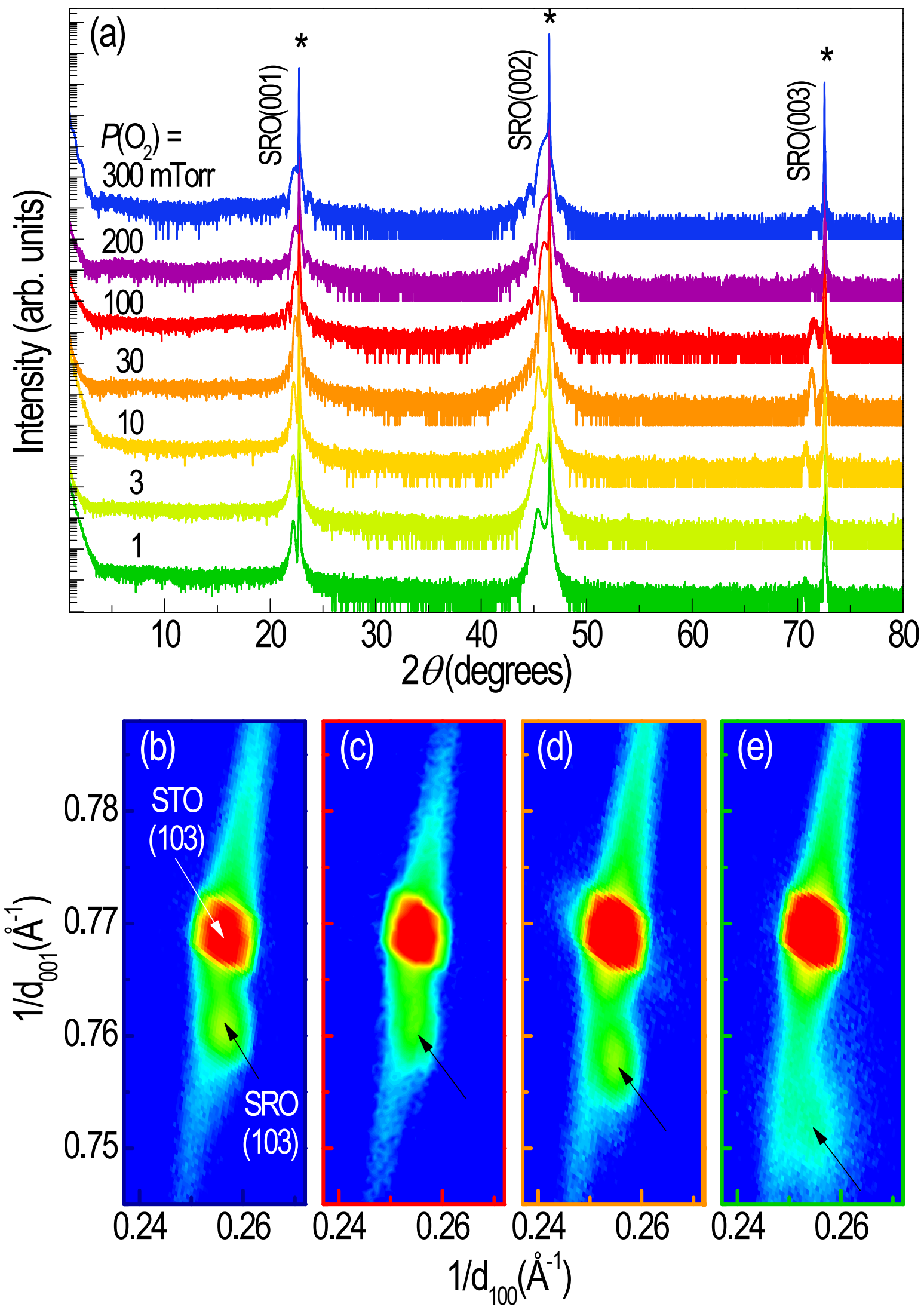

**Figure 1 | High quality of heteroepitaxial $SrRuO_3$ thin films with varying *P*($O_2$).** (a) XRD $\theta$-$2\theta$ scans for epitaxial $SrRuO_3$ thin films grown at different $P(O_2)$ around the (002) Bragg reflections of the $SrTiO_3$ substrates (*). With decreasing $P(O_2)$, the (002) peak of $SrRuO_3$ shifts to a lower angle, indicating an increase of the *c*-axis lattice constant. XRD reciprocal space mapping of the $SrRuO_3$ thin film grown at $P(O_2)$ = (b) 300, (c) 100, (d) 30, and (e) 1 mTorr around the (103) Bragg reflection of the $SrTiO_3$ substrate, which shows a coherently strained film with the same in-plane lattice constant, respectively.

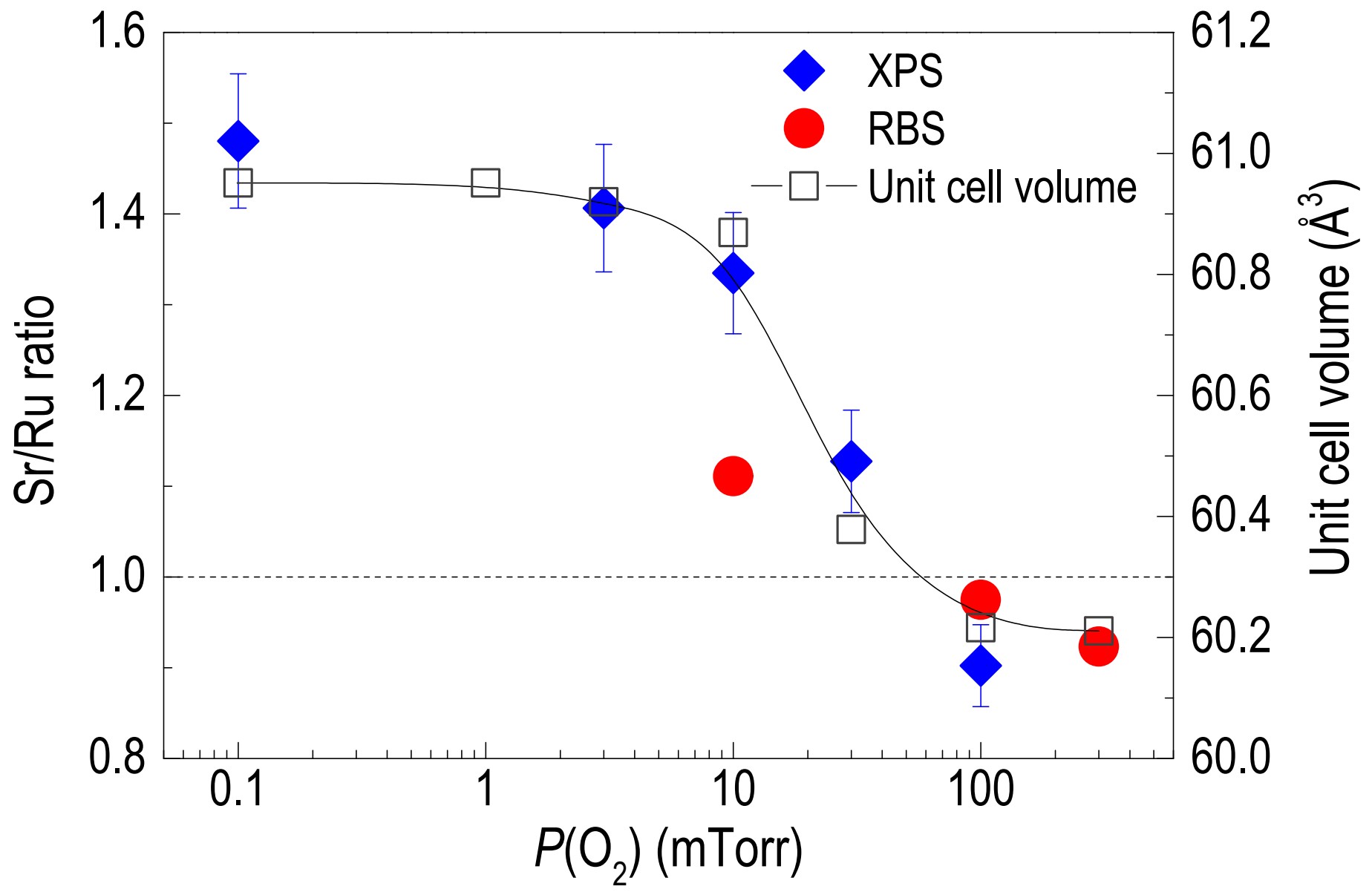

**Figure 2 | A comparison between the elemental defect concentration by x-ray photoemission spectroscopy and Rutherford back scattering spectroscopy and unit cell volume of $SrRuO_3$ thin films.** The $SrRuO_3$ thin film grown at $P(O_2)$ = 100 mTorr is stoichiometric, while Ru (Sr) deficiency is observed for the thin films grown below (above) $P(O_2)$ = 100 mTorr. The unit cell volume of the $SrRuO_3$ thin film is closely related to the cation stoichiometry.

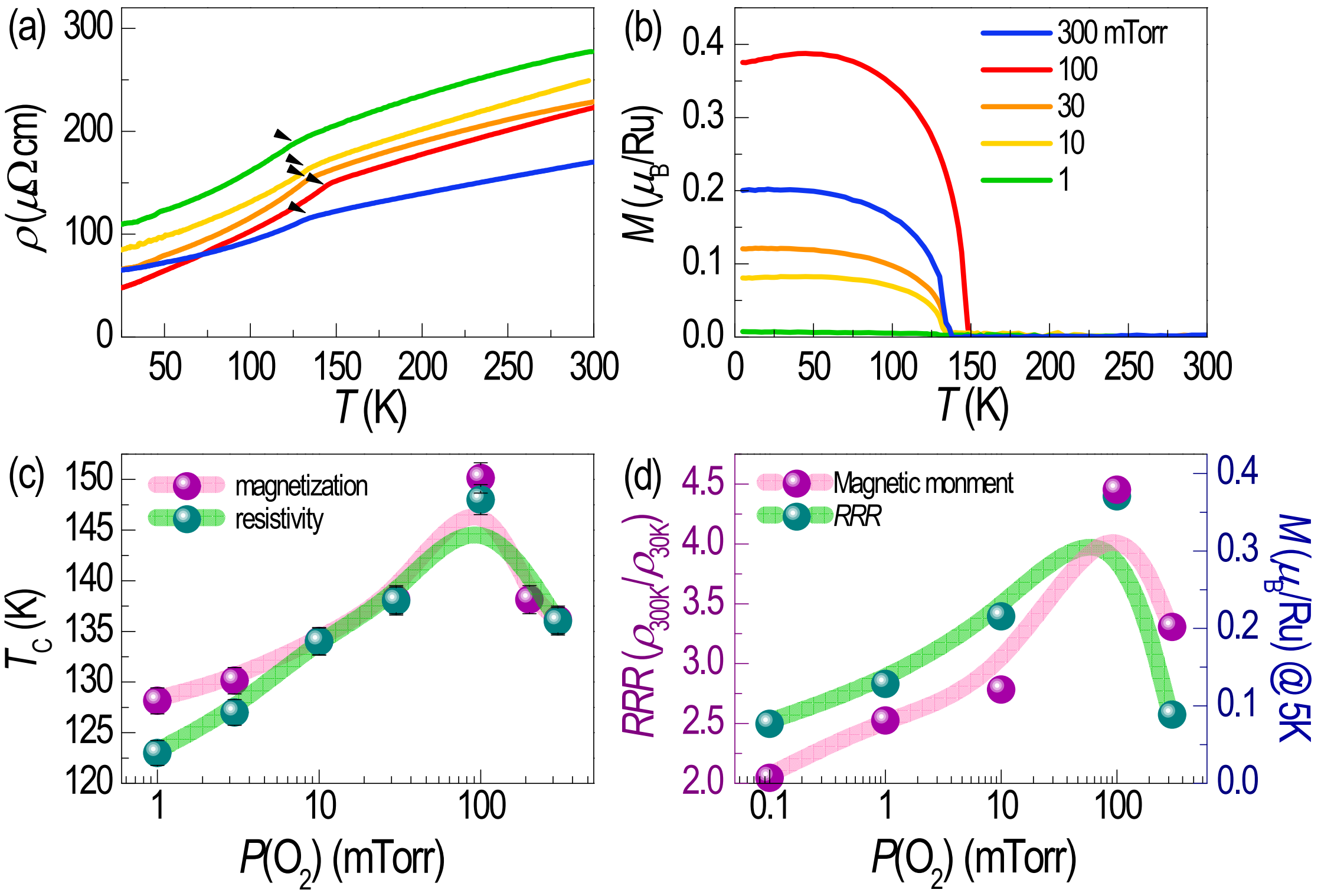

**Figure 3 | Changes in electrical and magnetic properties of $SrRuO_3$ thin films.** (a) Resistivity ($\rho(T)$) and (b) magnetic moment as a function of temperature for $SrRuO_3$ thin films deposited at different $P(O_2)$ at 10 Oe. (c) The $\rho(T)$ and $M(T)$ results show ferromagnetic transition temperature ($T_C$) below 150 K. (d) $P(O_2)$ dependence of the residual resistivity ratio ($RRR$) and the magnetization of the $SrRuO_3$ thin film. It decreasing under $P(O_2)$ = 100 mTorr, the $\rho(T)$ result shows clear anomaly, indicating the ferromagnetic ordering. The magnetization value measured at 5 K.

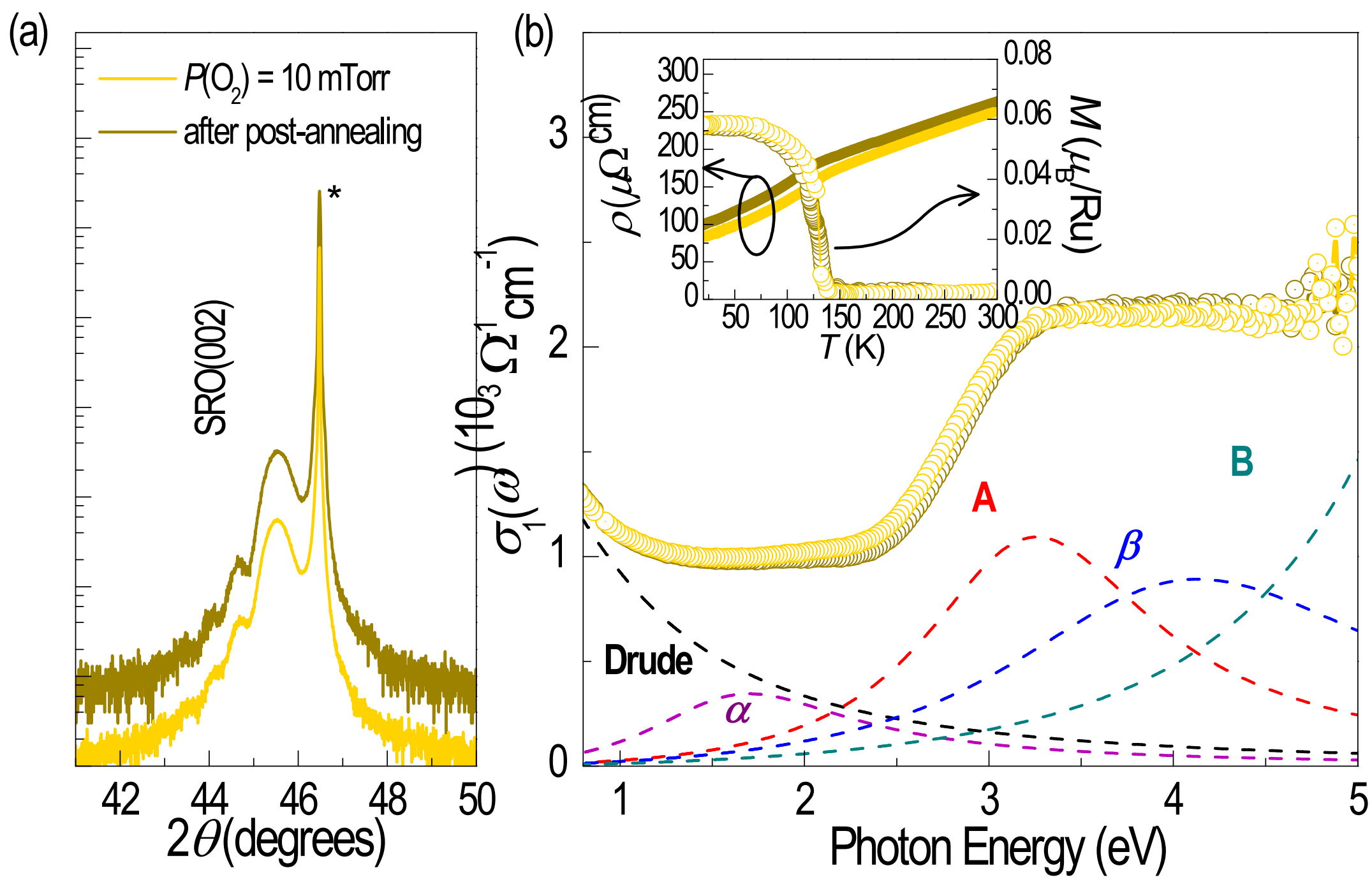

**Figure 4 | Crystal and electronic structure of $SrRuO_3$ thin films.** Robust (a) crystal and (b) optical properties of the $SrRuO_3$ thin films before and after thermal annealing at 700°C for 2 hours in air. The inset shows resistivity and magnetic moment under 100 Oe as a function of temperature for $SrRuO_3$ thin films before and after thermal annealing.

# Supplementary Information

# Tuning electromagnetic properties of $SrRuO_3$ epitaxial thin films via atomic control of cation vacancies

Sang A Lee[1], Seokjae Oh[1], Jegon Lee[1], Jae-Yeol Hwang[2], Jiwoong Kim[3], Sungkyun Park[3], Jong-Seong Bae[4], Tae Eun Hong[4], Suyoun Lee[5], Sung Wng Kim[2], Won Nam Kang[1], and Woo Seok Choi[1*]

[1]Department of Physics, Sungkyunkwan University, Suwon, 16419, Korea

[2]Department of Energy Sciences, Sungkyunkwan University, Suwon 16419, Korea

[3]Department of Physics, Pusan National University, Busan 46241, Korea

[4]Busan Center, Korea Basic Science Institute, Busan 46742, Korea

[5]Electronic Materials Research Center, Korea Institute of Science and Technology, Seoul 02792, Korea

[*]e-mail: choiws@skku.edu.

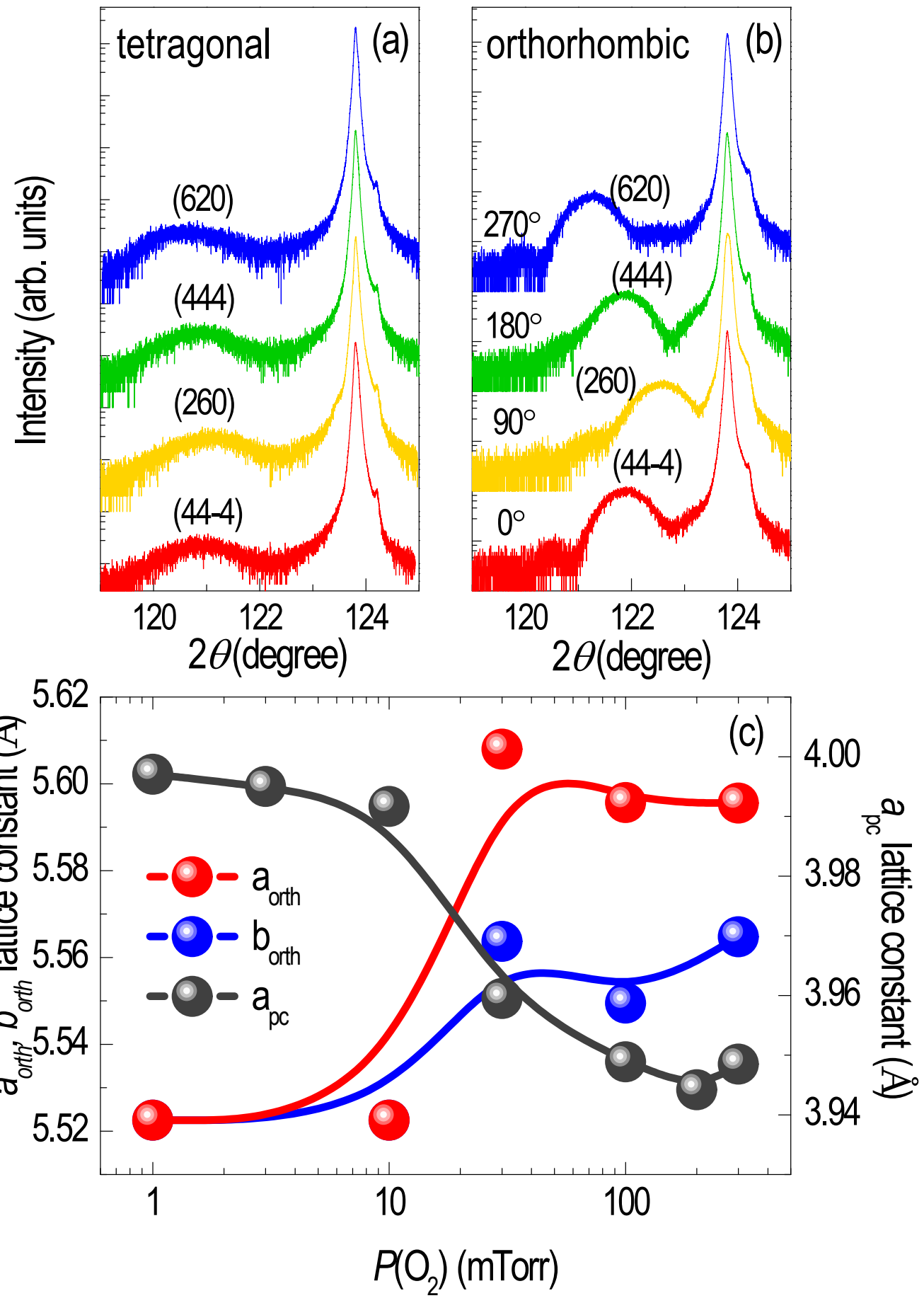

**Supplementary Figure S1.** Off-axis x-ray diffraction for the orthorhombic and tetragonal $SrRuO_3$ thin films. The $SrRuO_3$ thin films grown at at $P(O_2)$ = (a) 10 and (b) 100 mTorr, around the $SrTiO_3$ (204) Bragg reflections with configuration of $\varphi$ = 0, 90, 180, and 270°. (c) Evolution of the orthorhombic and pseudocubic lattice constants of epitaxial $SrRuO_3$ thin films as a function of $P(O_2)$.

| $P(O_2)$ (mTorr) | $a_o$ (Å) | $b_o$ (Å) | $c_o$ (Å) | $\gamma_o$ (°) | $\alpha_c$ (°) |
|---|---|---|---|---|---|
| 300 | 5.592 | 5.548 | 7.810 | 89.020 | 89.548 |
| 100 | 5.596 | 5.549 | 7.810 | 88.976 | 89.524 |
| 30 | 5.608 | 5.564 | 7.810 | 88.709 | 89.547 |
| 10 | 5.523 | 5.523 | 7.983 | 90 | 90 |
| 1 | 5.523 | 5.523 | 7.986 | 90 | 90 |

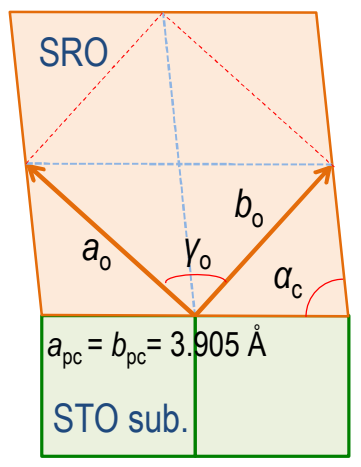

**Supplementary Table. S1. Lattice parameters of $SrRuO_3$ thin films grown at various $P(O_2)$.** Where $a_o$, $b_o$, $c_o$, $\gamma_o$, and $\alpha_c$ are the distorted orthorhombic unit-cell lengths, orthorhombic and pseudocubic tilt angle, respectively. The right side figure is unit cell of $SrRuO_3$ under compressive strain.

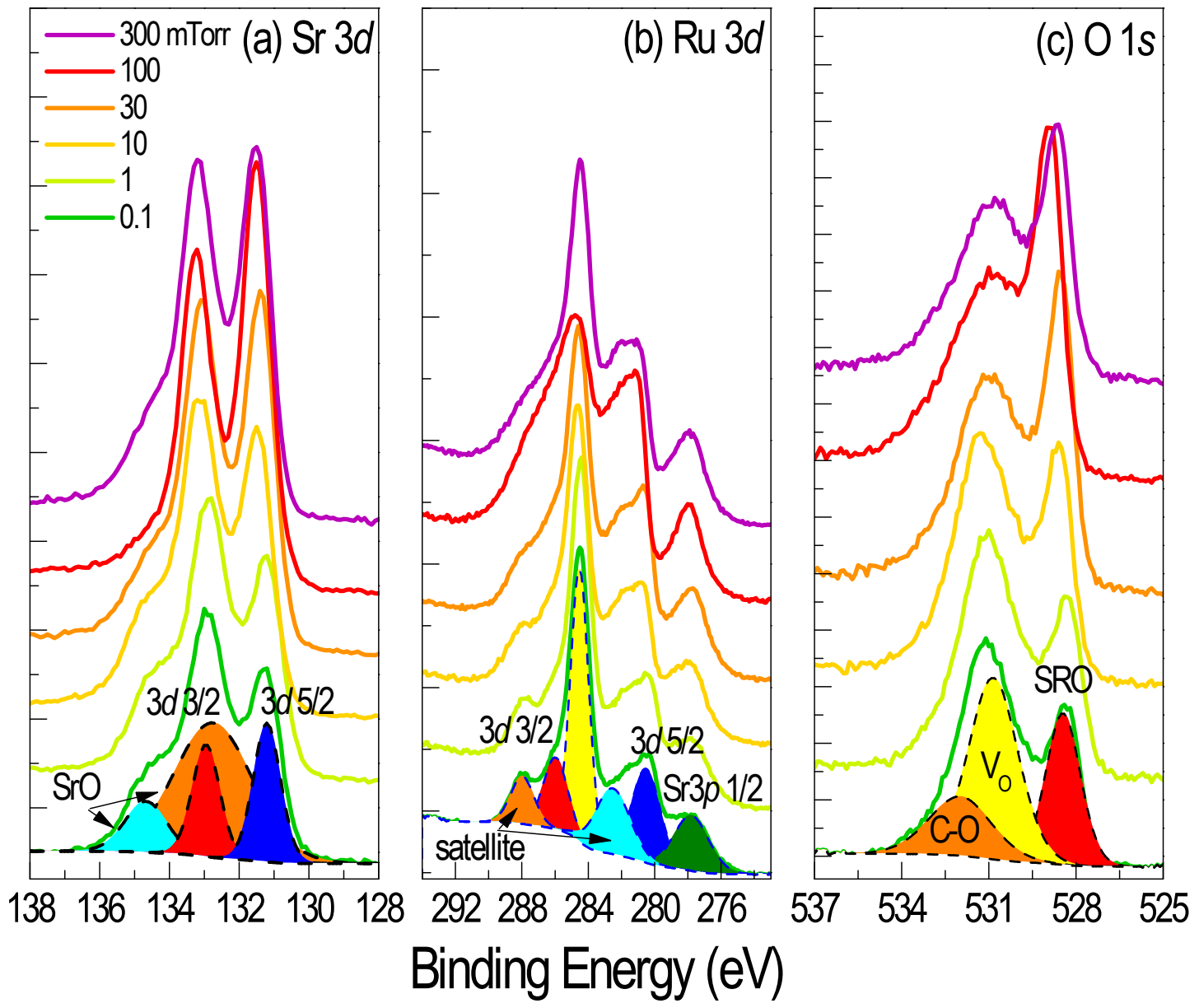

**Supplementary Figure S2.** X-ray photoemission spectroscopy (XPS) results of (a) Sr 3*d*, (b) Ru 3*d* core-level and (c) O 1*s* spectra for the $SrRuO_3$ thin films grown at different $P(O_2)$.

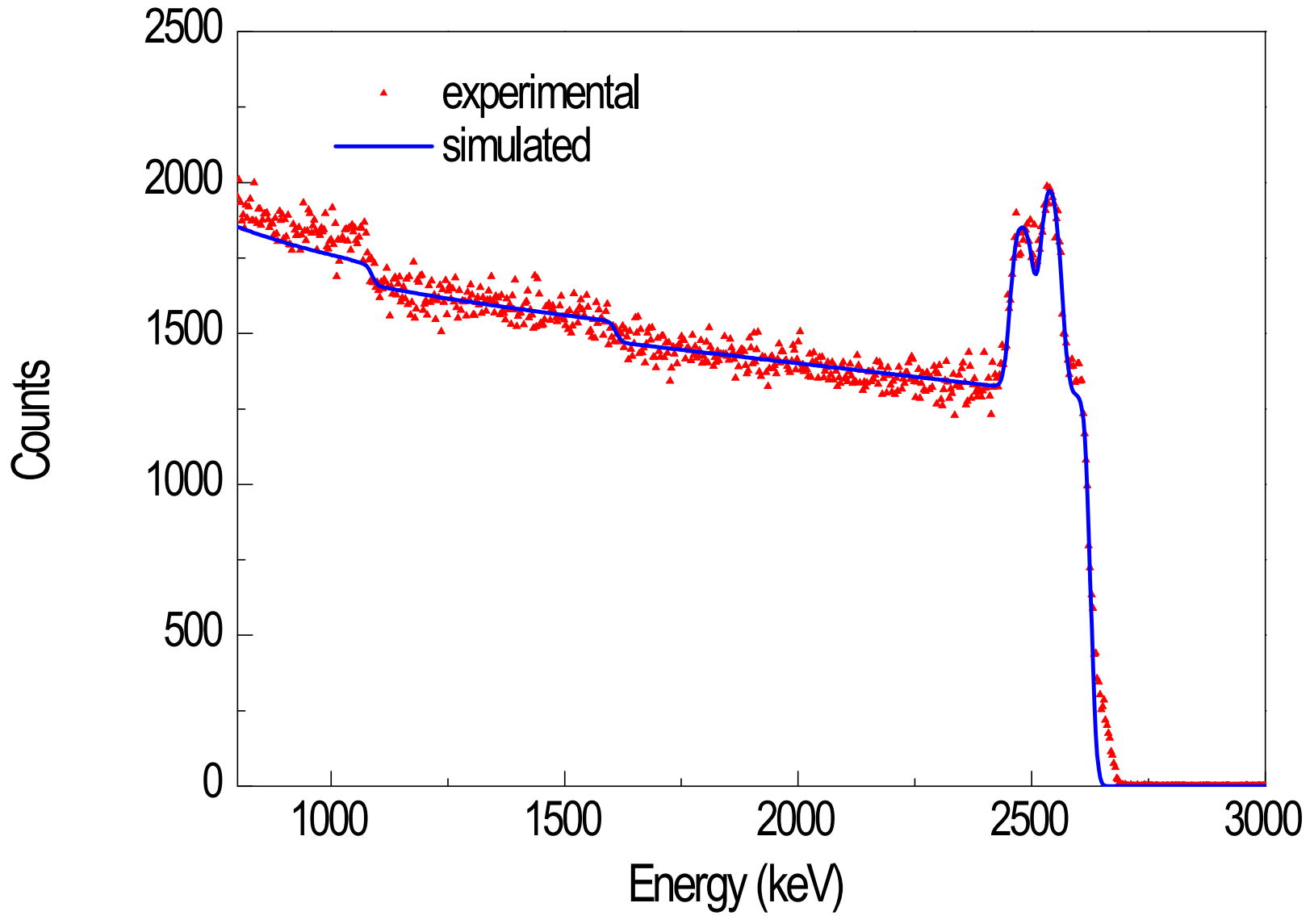

**Supplementary Figure S3.** The Rutherford backscattering (RBS) data of $SrRuO_3$ thin film.

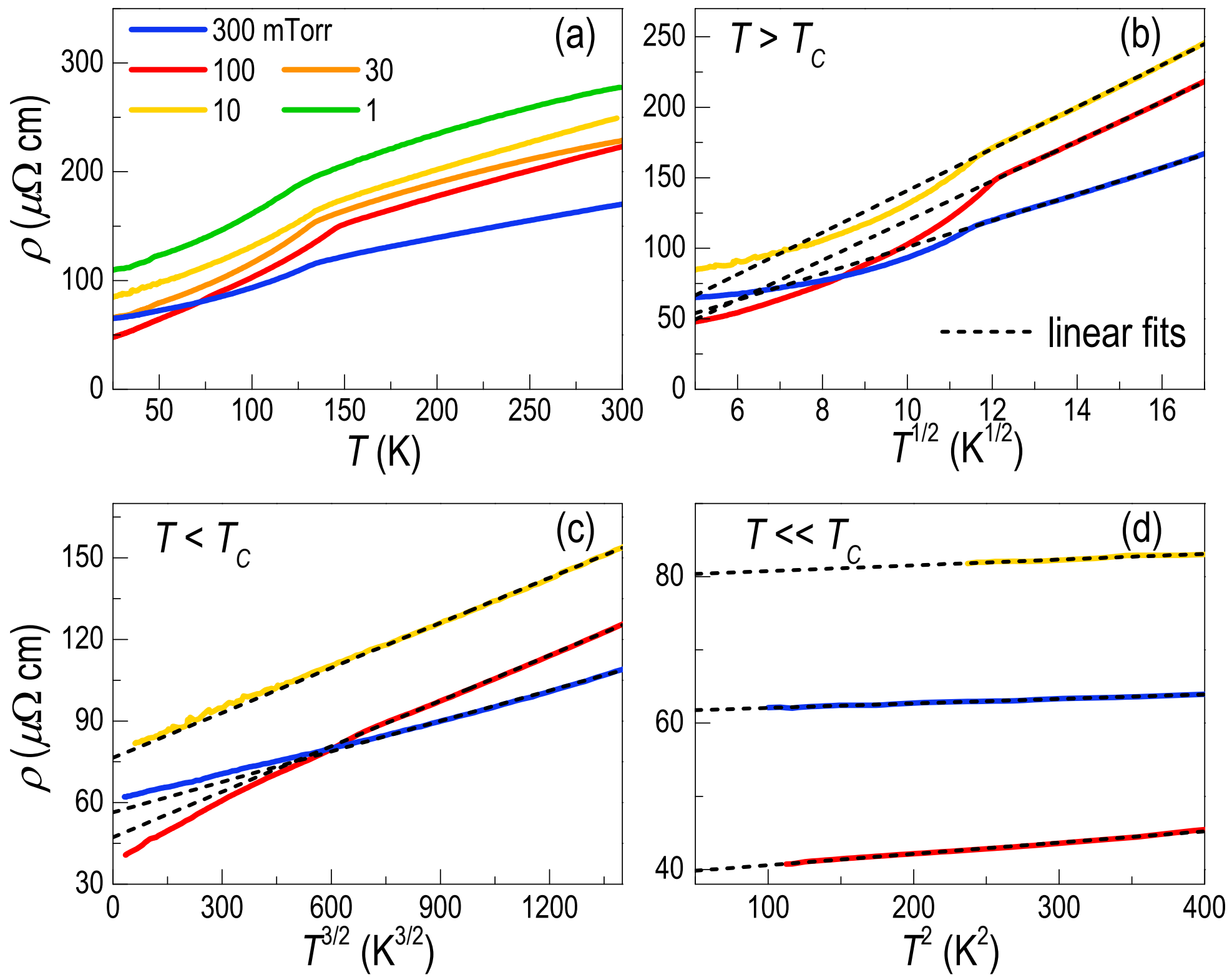

**Supplementary Figure S4.** Changes in electrical properties of $SrRuO_3$ thin films. (a) Resistivity as a function of temperature ($\rho(T)$) for $SrRuO_3$ thin films deposited at different $P(O_2)$. The resistivity of the $SrRuO_3$ thin films with different $P(O_2)$ are fitted with the power law $\rho(T) = \rho_0 + AT^{\alpha}$ ($\rho_0$ is residual resistivity, $A$ is temperature-dependent coefficient, $\alpha$ is scaling parameter, and $T$ is temperature) at (b) $T > T_C$, (c) $T < T_C$, and $T << T_C$.

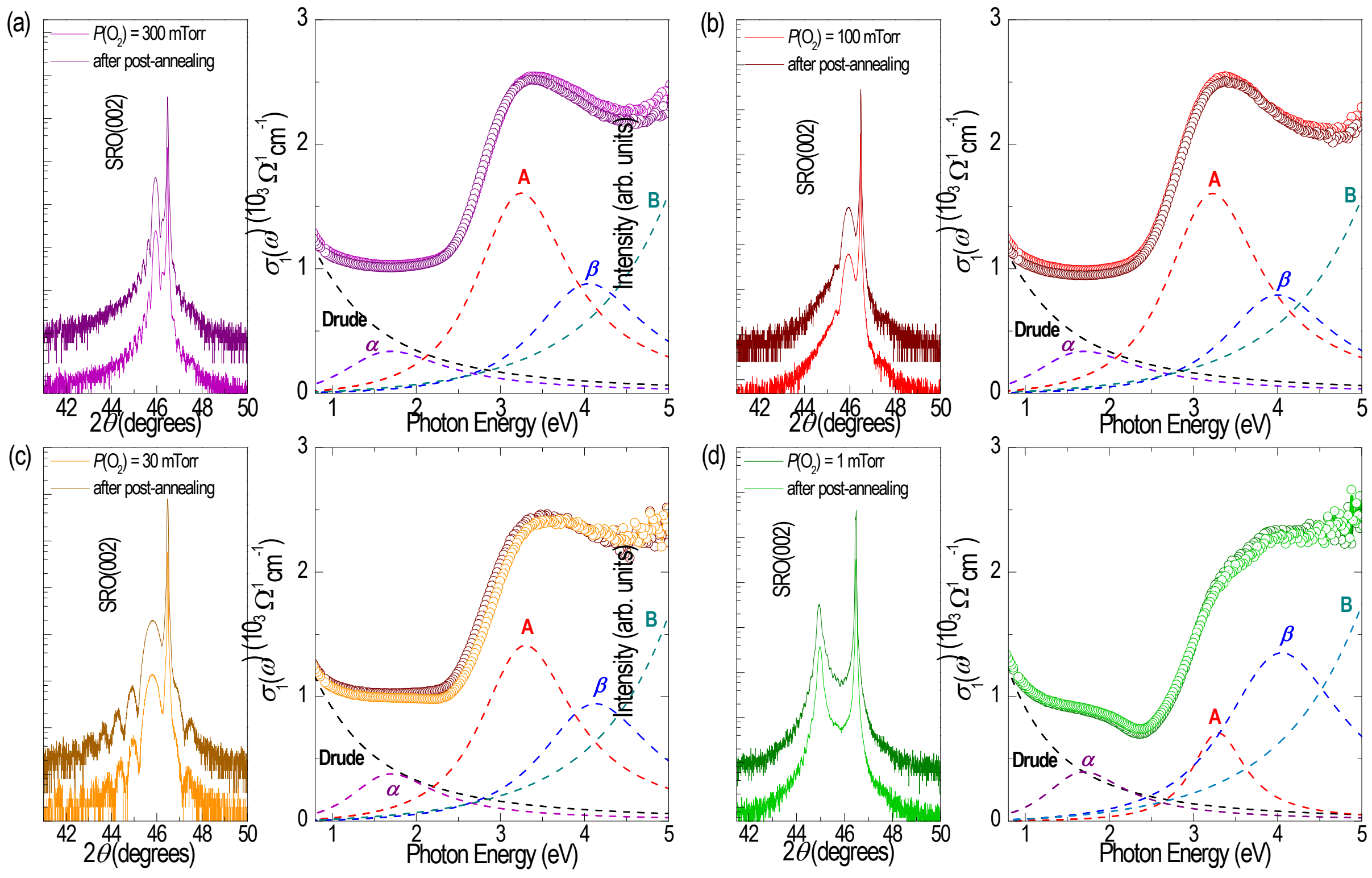

**Supplementary Figure S5.** Crystal and electronic structure of $SrRuO_3$ thin films. Robust crystal and optical properties of the $SrRuO_3$ thin films with $P(O_2)$ = (a) 300, (b) 100, (c) 30, and (d) 1 mTorr before and after thermal annealing at 700°C for 2 h in air.